\documentclass[epj]{svjour}
\usepackage{graphics}
\usepackage{amsmath,amssymb,bm,ascmac,calc}
\begin{document}
\title{Can realistic interaction be useful
for nuclear mean-field approaches?}
\author{H.~Nakada\inst{1} \and K.~Sugiura\inst{1}
\and T.~Inakura\inst{1,2,3} \and J.~Margueron\inst{4}
}                     
\offprints{}          
\institute{Department of Physics, Graduate School of Science,
 Chiba University,
Yayoi-cho 1-33, Inage, Chiba 263-8522, Japan
\and Yukawa Institute of Theoretical Physics, Kyoto University,
Kitashirakawa Oiwake-cho, Sakyo, Kyoto 606-8502, Japan
\and Department of Physics, Niigata University, Niigata 950-2181, Japan
\and Universit\'{e} de Lyon 1, CNRS/IN2P3,
 Institut de Physique Nucl\'{e}aire de Lyon, F-69622 Villeurbanne, France}
\date{Received: date / Revised version: date}
%
\abstract{
Recent applications of the M3Y-type semi-realistic interaction
to the nuclear mean-field approaches are presented:
(i) Prediction of magic numbers and
(ii) isotope shifts of nuclei with magic proton numbers.
The results exemplify that realistic interaction,
which is derived from the bare $2N$ and $3N$ interaction,
furnishes a new theoretical instrument
for advancing nuclear mean-field approaches.
} 
\maketitle
\section{Introduction}
\label{sec:intro}

Since atomic nuclei are self-bound systems,
self-consistent approaches are of particular importance
for the theoretical description of nuclear structure.
The self-consistent mean-field (MF)
or energy-density-functional (EDF) approaches,
\textit{i.e.} the Hartree-Fock (HF)
and Hartree-Fock-Bogolyubov (HFB) approaches,
are suitable for describing structure of nuclei
in a wide range of the nuclear mass table.
One of the great advantages of these approaches
is the self-consistency between the one-body fields
and the single-particle (s.p.) wave functions~\cite{ref:RS80a}.
In nuclear physics, the self-consistent HF calculations were started
by applying the zero-range Skyrme interaction~\cite{ref:VB72,ref:S3}.
While they have been successful,
the nuclear interaction is known to be short but finite range.
The self-consistent HFB calculations were first implemented
by Daniel Gogny and his collaborators~\cite{ref:Gog73,ref:Gogny}
with a finite-range interaction,
the so-called Gogny interactions.

Since it is comprised only of contact terms,
the HF energy derived from Skyrme-type interactions~\cite{ref:Sky,ref:Sky2}
is expressed in terms of the quasi-local EDF,
\textit{i.e.} by using only local currents
including derivatives of the s.p. functions.
Because of its computational simplicity
originating from its quasi-local character,
the Skyrme EDF has been most widely used in the self-consistent MF calculations.
While the HF potential includes only the particle-hole ($ph$) channel
of the effective interaction,
the particle-particle ($pp$) channel defines the pairing properties
and enters in the HFB equation.
The Skyrme EDF was later extended to include the $pp$ channel.
However, since a contact form of attraction is taken
also for the $pp$ channel,
certain cutoff scheme is required to avoid divergence.
The cutoff usually depends on the solution (\textit{e.g.} the s.p. energies),
and thereby violates the variational character
even though its influence could be small.
Moreover, the $pp$ channel is assumed to be independent of the $ph$ channel
in the Skyrme EDF, except the SkP parameter-set~\cite{ref:SkP}.
It is shown that the microscopic effective interaction
is different between the $ph$ and the $pp$ channels
in infinite systems such as nuclear matter~\cite{ref:RS80b},
because the s.p. space is separated between these channels.
This may hold also in vicinity of doubly magic nuclei.
In the density-functional theory (\textit{e.g.} the Kohn-Sham theory),
there is no need to assume underlying interaction,
and thereby no relation between the $ph$ and the $pp$ channels.
It is noted, however,
that the s.p. functions in the Kohn-Sham theory are artifacts,
losing their physical meaning.
In the HFB calculations for finite open-shell nuclei,
nucleons on certain orbitals can lie in the $ph$ and the $pp$ channels
simultaneously.
It will be natural that there is an interaction
consistent for both channels in that case,
and finite-range interactions offer the possibility to define
a unique interaction for these channels.

Owing to the Gaussian form for the central force,
the Gogny interaction is free of these problems.
The central force of the original parameter-set D1~\cite{ref:Gogny}
was partially related to the Brueckner HF results in the nuclear matter.
To some extent, it could be said that the D1 interaction has a nature
not far from the realistic nuclear interaction.
The connection to the realistic interaction
was lost in the phenomenologically `modified' parameter-sets
that came after such as D1S~\cite{ref:D1S} and D1M~\cite{ref:D1M}.
For the non-central part of these forces,
the contact interaction type is usually adopted.
This is the case for the LS force,
which is identical to the one of the Skyrme interaction.
The tensor force has been ignored
in the widely used parameter-sets of the Gogny interaction.
Although several parameter-sets including the tensor force
have been proposed~\cite{ref:OMA06,ref:ACD11,ref:CDAL12},
underlying the importance of the tensor force
in the MF regime~\cite{ref:Vtn,ref:SC14},
their application has yet been limited.

One of the authors (H.N.) has developed
another type of finite-range effective interaction
applicable to the self-consistent MF calculations;
the M3Y-type interaction~\cite{ref:Nak03,ref:Nak13}.
This may be categorized as \textit{semi-realistic}.
It is based on the M3Y interaction~\cite{ref:M3Y,ref:M3Y-P}
that was derived from the $G$-matrix,
and includes phenomenological modification
so as to reproduce the saturation and $\ell s$ splitting in finite nuclei.
Because of its connection to the realistic interaction,
the M3Y-type interaction will be suitable
to address the question raised as the title of this paper.
In this paper we shall mainly present results using the M3Y-P6 parameter-set
and its variant,
particularly focusing on the non-central forces.
M3Y-P$n$ with $n\geq 5$, including M3Y-P6, contains a tensor force
that is the same as the one
contained in the M3Y-Paris interaction~\cite{ref:M3Y-P}.
This enables us to study roles of the tensor force on a realistic basis.
An overall factor is applied to the LS force
in order to obtain reasonable s.p. level sequence in $^{208}$Pb.
For this part, a variant of M3Y-P6 has been introduced in Ref.~\cite{ref:NI15},
motivated by the recent finding
related to the three-nucleon ($3N$) interaction.
This new interaction reinforces the link to the realistic interaction.

\section{M3Y-type interaction and mean-field calculations}
\label{sec:int}

The effective Hamiltonian is $H=H_N+V_C-H_\mathrm{c.m.}$,
where $H_N$, $V_C$ and $H_\mathrm{c.m.}$\,($=\mathbf{P}^2/2AM$
with $\mathbf{P}=\sum_i\mathbf{p}_i$) represent
the nuclear Hamiltonian, the Coulomb interaction 
and the center-of-mass (c.m.) Hamiltonian, respectively.
The nuclear Hamiltonian $H_N$ is assumed to be
\begin{equation}\label{eq:H_N}
H_N = K + V_N\,;\quad K = \sum_i \frac{\mathbf{p}_i^2}{2M}\,,\quad
V_N = \sum_{i<j} v_{ij}\,,
\end{equation}
where $i$ and $j$ are the indices of individual nucleons.
The effective interaction $v_{ij}$ is built upon the following terms,
\begin{equation}\label{eq:effint1}
 v_{ij} = v_{ij}^{(\mathrm{C})}
 + v_{ij}^{(\mathrm{LS})} + v_{ij}^{(\mathrm{TN})}
 + v_{ij}^{(\mathrm{C}\rho)} + v_{ij}^{(\mathrm{LS}\rho)}\,,
\end{equation}
with
\begin{equation}\label{eq:effint2}
\begin{split}
v_{ij}^{(\mathrm{C})} &= \sum_n \big(t_n^{(\mathrm{SE})} P_\mathrm{SE}
+ t_n^{(\mathrm{TE})} P_\mathrm{TE} + t_n^{(\mathrm{SO})} P_\mathrm{SO} \\
&\hspace*{4cm} + t_n^{(\mathrm{TO})} P_\mathrm{TO}\big)
 f_n^{(\mathrm{C})} (r_{ij})\,,\\
v_{ij}^{(\mathrm{LS})} &= \sum_n \big(t_n^{(\mathrm{LSE})} P_\mathrm{TE}
 + t_n^{(\mathrm{LSO})} P_\mathrm{TO}\big)
 f_n^{(\mathrm{LS})} (r_{ij})\,\\
&\hspace*{5cm} \cdot[\mathbf{L}_{ij}\cdot(\mathbf{s}_i+\mathbf{s}_j)]\,,\\
v_{ij}^{(\mathrm{TN})} &= \sum_n \big(t_n^{(\mathrm{TNE})} P_\mathrm{TE}
 + t_n^{(\mathrm{TNO})} P_\mathrm{TO}\big)
 f_n^{(\mathrm{TN})} (r_{ij})\, r_{ij}^2 S_{ij}\,,\\
v_{ij}^{(\mathrm{C}\rho)} &= \big(C^{(\mathrm{SE})}[\rho(\mathbf{r}_i)]\,P_\mathrm{SE}
 + C^{(\mathrm{TE})}[\rho(\mathbf{r}_i)]\,P_\mathrm{TE}\big)
 \,\delta(\mathbf{r}_{ij})\,.
\end{split}
\end{equation}
Here $\mathbf{s}$ is the spin operator,
$\mathbf{r}_{ij}= \mathbf{r}_i - \mathbf{r}_j$,
$\mathbf{p}_{ij}= (\mathbf{p}_i - \mathbf{p}_j)/2$,
$\mathbf{L}_{ij}= \mathbf{r}_{ij}\times \mathbf{p}_{ij}$,
$S_{ij}= 4\,[3(\mathbf{s}_i\cdot\hat{\mathbf{r}}_{ij})
(\mathbf{s}_j\cdot\hat{\mathbf{r}}_{ij})
- \mathbf{s}_i\cdot\mathbf{s}_j ]$
and $\rho(\mathbf{r})$ denotes the nucleon density.
$P_\mathrm{Y}$ denotes the projection operator
on the two-particle channel $\mathrm{Y}$
($\mathrm{Y}=\mathrm{SE},\mathrm{TE},\mathrm{SO},\mathrm{TO}$).
The interaction $v_{ij}^{(\mathrm{LS}\rho)}$ is used in Sec.~\ref{sec:shift},
and its form will be shown there.

The Skyrme interaction~\cite{ref:Sky} assumes zero-range functions
$\delta(\mathbf{r}_{ij})$ or $\nabla^2\delta(\mathbf{r}_{ij})$
for the form factor $f_n^{(\mathrm{X})}(r)$ ($\mathrm{X}
=\mathrm{C},\,\mathrm{LS},\,\mathrm{TN}$),
generating a quasi-local form for the resultant EDF.
For the Gogny interaction,
$f_n^{(\mathrm{C})}(r)$ is taken to be Gaussian $e^{-(\mu_n r)^2}$,
while $f^{(\mathrm{LS})}(r)$ is set to be $\nabla^2\delta(\mathbf{r})$
and the tensor interaction $v^{(\mathrm{TN})}$ is ignored
in the widely used parameter-sets.
For the M3Y-type interactions
we take $f_n^{(\mathrm{X})}(r)$ to be the Yukawa function $e^{-\mu_n r}/\mu_n r$
for all of the density-independent channels
$\mathrm{X}=\mathrm{C},\,\mathrm{LS},\,\mathrm{TN}$.
As in the Skyrme and Gogny interactions,
a density-dependent contact term $v^{(\mathrm{C}\rho)}$ is introduced,
so as to reproduce the saturation properties,
with $C^{(\mathrm{Y})}[\rho]=t_\rho^{(\mathrm{Y})}\rho^{\alpha^{(\mathrm{Y})}}$.
As mentioned earlier, $v^{(\mathrm{TN})}$ in M3Y-P6 is identical
to that in the M3Y-Paris interaction~\cite{ref:M3Y-P}.

Concerning numerical calculations,
the methods detailed in Ref.~\cite{ref:Nak06} is applied
with the basis functions given in Ref.~\cite{ref:Nak08}.
The exchange and the pairing terms of $V_C$ are explicitly taken into account,
as well as the $2$-body term of $H_\mathrm{c.m.}$.

\section{Prediction of magic numbers}
\label{sec:magic}

Magic numbers are manifestation of the shell structure,
which is one of fundamental ingredients of the nuclear structure theory.
While magic numbers near the $\beta$ stability are well established,
experiments using radioactive beams disclosed
that magic numbers may appear and disappear far off the $\beta$ stability,
depending on $Z$ and $N$ numbers~\cite{ref:SP08}.
The $Z$- and $N$-dependence of the shell structure
is sometimes called \textit{shell evolution}~\cite{ref:Ots08}.
It has been pointed out that the tensor force $v^{(\mathrm{TN})}$
plays important roles
in the shell evolution~\cite{ref:Vtn,ref:Nak10b,ref:NSM13}.
It has also been argued
that $\ell$-dependence of the loosely bound s.p. levels
could contribute to the shell evolution~\cite{ref:N16}.
The self-consistent MF approaches with M3Y-type semi-realistic interaction,
which includes the realistic tensor force,
are suitable for investigating magic numbers
in the wide range of the nuclear mass table
going from stable to unstable nuclei.
It is noted, however, that loosely bound orbitals shall be properly handled
with appropriate numerical methods~\cite{ref:Nak06,ref:Nak08,ref:NS02}.

\subsection{$p0d_{3/2}$-$p1s_{1/2}$ inversion in Ca isotopes}
\label{subsec:Ca-inv}

Although the s.p. levels are difficult to be observed unambiguously
even for doubly magic nuclei,
because of their fragmentation due to core polarization,
they could experimentally be extracted
by taking averaged energies weighted by the spectroscopic factors.
Although there are not many such cases,
the spectroscopic factors of $p0d_{3/2}^{-1}$ and $p1s_{1/2}^{-1}$ levels
on top of $^{40}$Ca and $^{48}$Ca are almost exhausted
in experiments~\cite{ref:DWKM76,ref:Ogi87}.
Interestingly, the experimental results indicate
an inversion of these two levels from $^{40}$Ca to $^{48}$Ca.

For the doubly magic nuclei $^{40}$Ca and $^{48}$Ca,
the spherical HF is expected to give a good approximation
of the ground-state wave functions.
Then the $p0d_{3/2}$ and $p1s_{1/2}$ energies in the HF
can be compared to the experimental energies
averaged by the spectroscopic factors.
Some effective interactions reproduce
the $p1s_{1/2}$-$p0d_{3/2}$ inversion,
but others do not~\cite{ref:GMK07,ref:WGZD11}.
However, even with the interactions that reproduce the inversion,
the slope of the s.p. level spacing
$\mathit{\Delta}\varepsilon_{13}
=\varepsilon(p1s_{1/2})-\varepsilon(p0d_{3/2})$
from $^{40}$Ca to $^{48}$Ca is not reproduced,
except for the relativistic HF calculation with PKA1~\cite{ref:LMLG15}.

\begin{figure}
\resizebox{0.5\textwidth}{!}{%
  \includegraphics{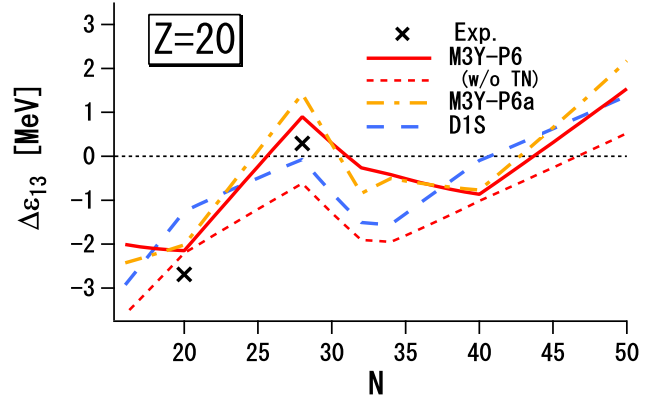}
}
\caption{$\mathit{\Delta}\varepsilon_{13}
=\varepsilon(p1s_{1/2})-\varepsilon(p0d_{3/2})$ for Ca isotopes.
Red solid, orange dot-dashed and blue long dashed lines represent
the spherical HF results with the M3Y-P6, M3Y-P6a and D1S interactions,
respectively.
Experimental data are obtained after average
weighted by the spectroscopic factors~\protect\cite{ref:DWKM76,ref:Ogi87}.
Thin red short dashed line is obtained from M3Y-P6
but by removing contribution of the tensor force
(see text for more details).}
\label{fig:dspe_Z20}       
\end{figure}

The s.p. energies are calculated for the individual Ca isotopes
within the spherical HF regime,
on the equal-filling assumption when necessary.
The slope is also quite well reproduced
with the semi-realistic M3Y-P$n$ interactions
as shown in Fig.~\ref{fig:dspe_Z20} and in Ref~\cite{ref:NSM13}.
Moreover, we find that the $N$-dependence of $\mathit{\Delta}\varepsilon_{13}$
is quite different from the $N$-dependence in the previous results
without tensor force,
as exemplified by the difference between the M3Y-P6 result
and the result of the Gogny-D1S interaction in Fig.~\ref{fig:dspe_Z20}.
To confirm the important role of the tensor-force,
we evaluate contribution of $v^{(\mathrm{TN})}$ to the s.p. energy as
\begin{equation}\label{eq:spe-TN}
\begin{split}
 \varepsilon^{(\mathrm{TN})}(j)
 &= \sum_{j'm'} n_{j'}\langle jmj'm'|v^{(\mathrm{TN})}|jmj'm'\rangle \\
 &= \frac{1}{2j+1}\sum_{j'J} n_{j'}(2J+1)\langle jj'J|v^{(\mathrm{TN})}|jj'J\rangle\,,
\end{split}
\end{equation}
where $n_{j'}$ denotes the occupation probability of the s.p. state $j'$.
If we subtract the contribution of $v^{(\mathrm{TN})}$
from the energy difference $\mathit{\Delta}\varepsilon_{13}$ with M3Y-P6,
defining it as $\varepsilon^{(\mathrm{TN})}(p1s_{1/2})
-\varepsilon^{(\mathrm{TN})}(p0d_{3/2})$,
the $N$-dependence of $\mathit{\Delta}\varepsilon_{13}$
becomes very similar to that of the D1S interaction (with a constant shift).
This shift of the absolute value is attributed to the central and LS forces,
which are adjusted under the presence (absence) of the tensor force
in M3Y-P6 (D1S).
These results confirm the importance of the tensor force,
and show that the realistic tensor force is useful
for describing the shell evolution.
Similar conclusions could also be found
in relativistic HF approaches~\cite{ref:LMLG15}.

$\mathit{\Delta}\varepsilon_{13}$ calculated with M3Y-P5$'$, M3Y-P7
and D1M have been shown in Ref.~\cite{ref:NSM13}.
The absolute values of $\mathit{\Delta}\varepsilon_{13}$
depend on the interactions to certain extent.
For instance, the $p1s_{1/2}$-$p0d_{3/2}$ inversion at $^{48}$Ca
is well reproduced with D1M, unlike D1S.
However, $N$-dependence of $\mathit{\Delta}\varepsilon_{13}$
is primarily determined by the tensor force as pointed out above.
The result of D1S nearly matches that of D1M if shifted by a constant,
as in the comparison of the D1S result
and the tensor-subtracted result of M3Y-P6.
The same situation holds among the M3Y-P$n$ interactions.
In Fig.~\ref{fig:dspe_Z20}, the result of M3Y-P6a,
which is different from M3Y-P6 only in the LS channel
and will be discussed in Sec.~\ref{sec:shift},
is also presented.
We observe that M3Y-P6a provides similar $N$-dependence
of $\mathit{\Delta}\varepsilon_{13}$ to M3Y-P6,
apart from the slope in $N\leq 20$ and the small staggering around $N=32$.

\subsection{Magic numbers}
\label{subsec:magic}

\begin{figure*}
\resizebox{1.0\textwidth}{!}{%
  \includegraphics{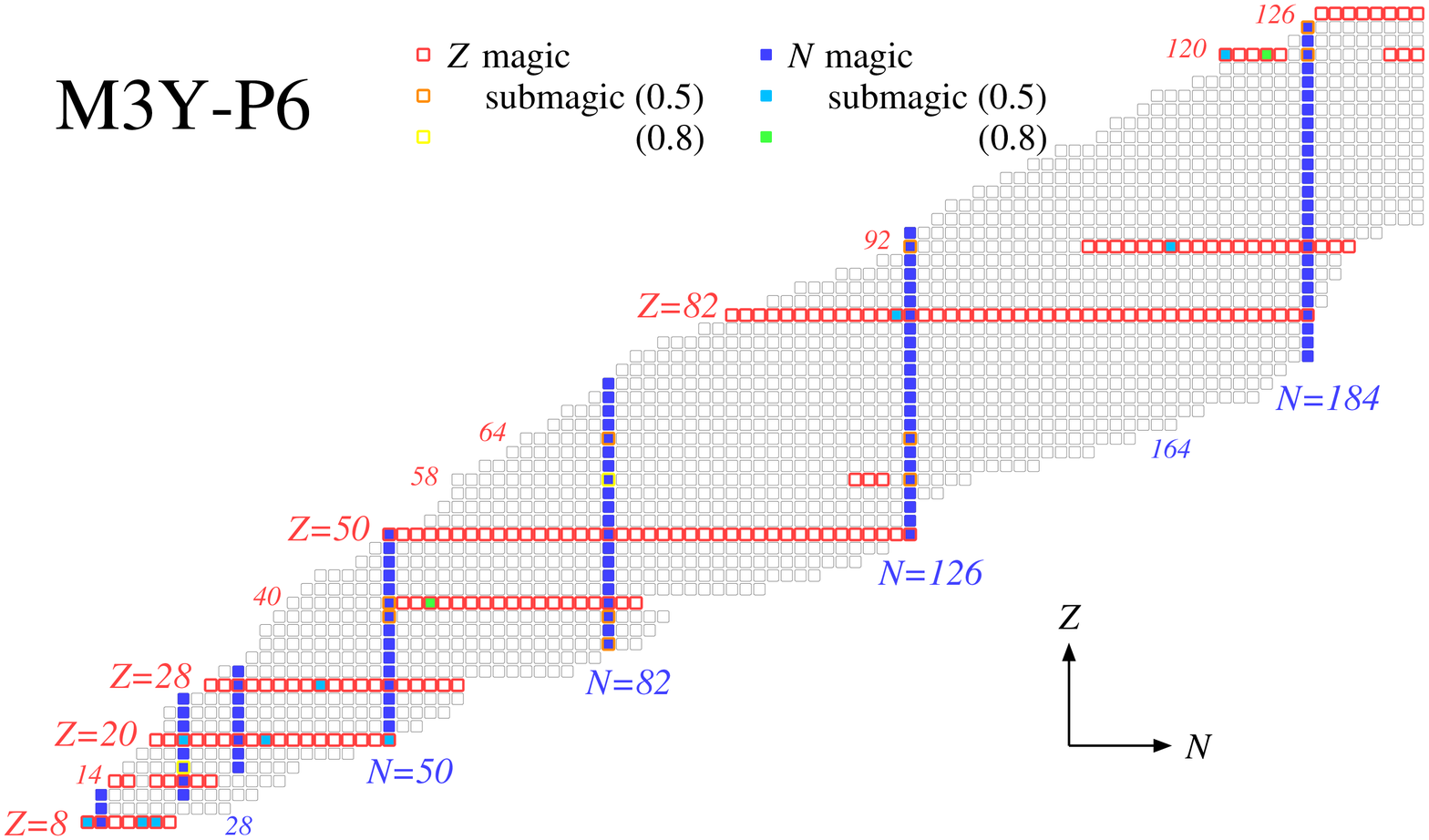}
}
\caption{Chart showing magic numbers predicted with the M3Y-P6 interaction.
Individual boxes correspond to even-even nuclei.
Magic (submagic) $Z$'s are represented
by red (orange and yellow) contour of boxes,
while magic (submagic) $N$'s by blue (skyblue and green) filled boxes.
For submagic nuclei, the values for $\lambda_\mathrm{sub}$
are indicated in parenthesis in the caption (in MeV).
Quote from Ref.~\protect\cite{ref:NS14}.}
\label{fig:magic}       
\end{figure*}

We have applied the M3Y-type semi-realistic interaction
to prediction of magic numbers in a wide range of the nuclear mass table.
From experimental viewpoints,
magic numbers have been identified
by the relative stability of certain nuclei
with respect to masses, excitation energies, and so forth.
However, there is no clear theoretical definition of magic numbers.
We first discuss how magic numbers are identified in this work.

Many-body correlations are strongly quenched at magicity,
because of large shell gap.
Spherical HF solution is therefore expected to provide a good approximation
for doubly magic nuclei,
as in the case of $^{40,48}$Ca discussed in the previous subsection.
It is not easy to quantitatively compare the size of the shell gap
to the strength of the many-body correlations.
A rather accurate theoretical measure of the many-body correlation
may however be given by pairing correlations,
since it is the main correlation beyond HF in spherical nuclei.
For practical identification of magic nuclei
we implement spherical HFB calculations,
and identify magic number in $Z$ ($N$)
when the proton (neutron) pair correlation vanishes.
In addition, \textit{submagic} numbers are identified
if the pair correlation is quite reduced.
Such suppression has been observed
particularly when either $Z$ or $N$ is already a good magic number,
drawing the counterpart to be submagic;
\textit{e.g.} $N=40$ at $^{68}$Ni~\cite{ref:Nak10b}
and $Z=64$ at $^{146}$Gd~\cite{ref:MNOM}.
We compare the HF and the HFB energies
(denoted by $E_\mathrm{HF}$ and $E_\mathrm{HFB}$),
and identify submagic $Z$ ($N$)
if the condensation energy $E_\mathrm{cond}=E_\mathrm{HF}-E_\mathrm{HFB}$
is smaller than a certain value $\lambda_\mathrm{sub}$
for $N=\mbox{magic}$ ($Z=\mbox{magic}$) nuclei.
The $\lambda_\mathrm{sub}$ value is rather arbitrary,
and we shall show results of $\lambda_\mathrm{sub}=0.5\,\mathrm{MeV}$
and $0.8\,\mathrm{MeV}$ below.
Considering the domain where MF approaches are rather valid
and the experimental accessibilities,
we restrict our calculations to $8\leq Z\leq 126$, $N\leq 200$.

When a nucleus is identified as not being magic,
it reveals a small shell gap that could not quench many-body correlations
in the present spherical HFB calculation.
Quadrupole deformation, which is not considered here,
could still be important in some cases.
The present criterion remains however quite useful
to pick up candidates for magicity.
Quadrupole deformation will be investigated in future studies.

Magic and submagic numbers predicted with the M3Y-P6 interaction
are shown in Fig.~\ref{fig:magic}.
It is noticeable that the results with M3Y-P6 are not contradictory
to available experimental data except in a few cases.
For instance, $N=16$ comes submagic at $^{24}$O, and so does
$N=32$ at $^{52}$Ca, $N=40$ at $^{68}$Ni, $Z=64$ at $^{146}$Gd.
The $N=28$ magicity is lost in $Z\leq 14$ while kept in $Z\geq 16$.
Although the disappearance of the $N=20$ magic number is correctly
reproduced for $^{30}$Ne,
it still is magic at $^{32}$Mg in Fig.~\ref{fig:magic},
which is one of the few exceptions.
This disagreement with the data will probably be reduced
by considering quadrupole deformation~\cite{ref:SNM16}.

In Ref.~\cite{ref:NS14}, more results are presented,
which are obtained from other interactions such as M3Y-P7 and D1M.
Unlike for the M3Y-P6 interaction,
predictions based on M3Y-P7 and D1M are much less consistent
with experimental data for a large number of nuclides.
A part of the success of the M3Y-P6 interaction
is certainly attributed to the contribution of the realistic tensor force,
although the other channels are also important
as the comparison to M3Y-P7 illustrates.

\begin{figure*}
\centerline{\resizebox{0.7\textwidth}{!}{%
  \includegraphics{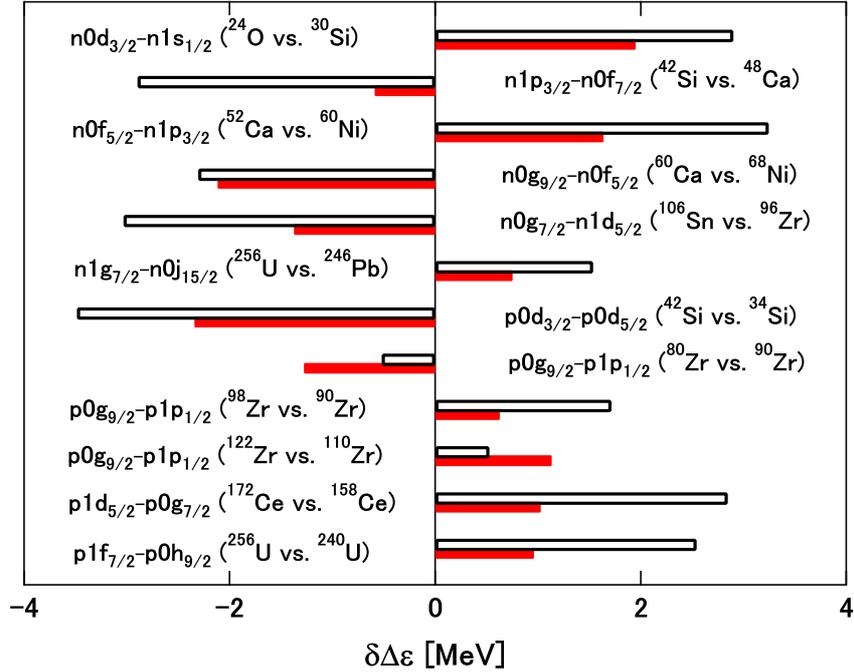}
}}
\caption{Difference of the shell gaps
 between two members of isotopes or isotones
 $\delta\mathit{\Delta}\varepsilon(j_2\,\mbox{-}\,j_1)$ (open bars),
 with contributions of $v^{(\mathrm{TN})}$ (red bars) to it,
 obtained from the HF results with M3Y-P6.}
\label{fig:ddspe_M3Yp6}       
\end{figure*}

To measure the effects of the tensor force on the shell gap,
the double difference of the s.p. energies,
$\mathit{\Delta}\varepsilon(j_2\,\mbox{-}\,j_1)
=\varepsilon(j_2)-\varepsilon(j_1)$ at a certain nuclide $(Z_b,N_b)$
relative to that at a reference nuclide $(Z_a,N_a)$,
is denoted by $\delta\mathit{\Delta}\varepsilon(j_2\,\mbox{-}\,j_1)$,
and is calculated for the HF results with M3Y-P6.
Replacing $\varepsilon(j)$ by $\varepsilon^{(\mathrm{TN})}(j)$
of Eq.~(\ref{eq:spe-TN}),
we also evaluate tensor-force contribution
to $\delta\mathit{\Delta}\varepsilon(j_2\,\mbox{-}\,j_1)$.
Some results are presented in Fig.~\ref{fig:ddspe_M3Yp6}.
For instance, the top row indicates
that the shell gap between $n0d_{3/2}$ and $n1s_{1/2}$ increases
by about $3\,\mbox{MeV}$ from $^{30}$Si to $^{24}$O,
and about $2\,\mbox{MeV}$ out of $3\,\mbox{MeV}$
comes from the tensor force.
This tells us importance of the tensor force
in the new magicity $N=16$ around $^{24}$O.
Significance of the tensor force is found also
in the $N=32$ magicity at $^{52}$Ca.
It is interesting to notice that similar results have been obtained
within a relativistic framework
including Lorentz tensor coupling~\cite{ref:LMLG15}.

\section{Isotope shifts for nuclei with magic proton numbers}
\label{sec:shift}

Accurate measurements of isotope shifts~\cite{ref:AHS87,ref:Ang04}
have questioned the predictivity of nuclear structure theories
for a long period.
One is the conspicuous kink at $N=126$ in the neutron-rich Pb isotope.
Although the kink itself can be reproduced
in a relativistic MF approach~\cite{ref:SLR94}
and in a non-relativistic approach with the modified Skyrme EDF~\cite{ref:RF95},
in those results the s.p. level spacing between $n1g_{9/2}$ and $n0i_{11/2}$
is found to be too small.
Another long-standing challenge for the MF theories is
the reproduction of the very close charge radii between $^{40}$Ca and $^{48}$Ca.
Both of these two nuclei are doubly magic
and hence expected to be well described by MF theories.
However, to our best knowledge, no self-consistent calculations
have been able to reproduce their close charge radii so far.

%
\begin{figure*}
\centerline{\resizebox{0.75\textwidth}{!}{%
  \includegraphics{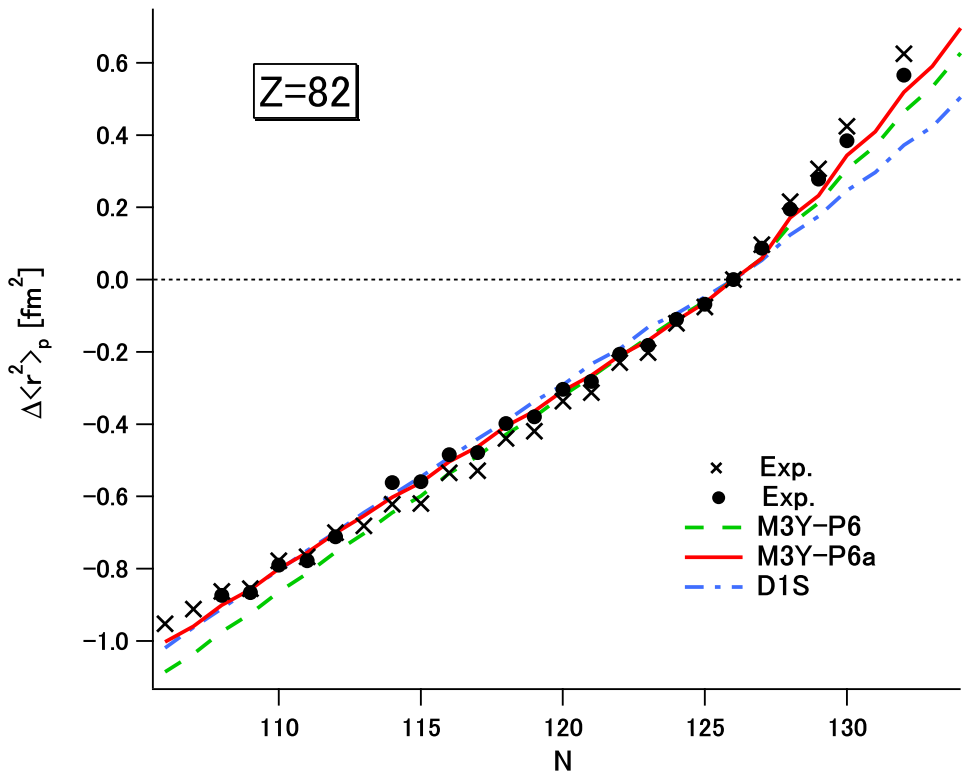}
}}
\caption{Isotope shifts of the Pb nuclei
$\mathit{\Delta}\langle r^2\rangle_p(\mbox{$^A$Pb})$,
obtained from the HFB calculations with M3Y-P6a (red solid line),
in comparison to those with M3Y-P6 (green dashed line)
and D1S (blue dot-dashed line).
Experimental data are quoted from Refs.~\protect\cite{ref:Ang13} (crosses)
and \protect\cite{ref:AHS87} (circles).}
\label{fig:Pb_drp}       
\end{figure*}

As shown below, these deficiencies are related to the LS interaction.
While the $\ell s$ splitting plays an essential role
in the nuclear shell structure,
it has been difficult so far to account for the observed size
of the $\ell s$ splitting
only from the $2N$ interaction~\cite{ref:AB81}.
Recently, Kohno has pointed out,
based on the chiral effective-field theory ($\chi$EFT),
that the $3N$ LS interaction may account for the missing part
of the $\ell s$ splitting~\cite{ref:Koh12,ref:Koh13}.
In the following, we explore this idea
and incorporate the $3N$ LS interaction in our MF model,
reinvestigating the issue of the isotope shifts of the nuclei
with magic proton numbers.

As shown in the previous section,
the M3Y-P6 interaction gives reasonable shell structure.
However, $v^{(\mathrm{LS})}$ in M3Y-P6 is not realistic,
since it is obtained phenomenologically
by multiplying $v^{(\mathrm{LS})}$ of the M3Y-Paris interaction
by an overall factor of about $2$ ($2.2$, to be precise).
Here we modify the LS channel of the M3Y-P6 interaction
by adding the $3N$ LS interaction suggested by Kohno,
instead of enhancing $v^{(\mathrm{LS})}$.
In practice we use a density-dependent LS interaction $v^{(\mathrm{LS}\rho)}$,
which is defined as
\begin{equation}
v_{ij}^{(\mathrm{LS}\rho)} = 2i\,D[\rho(\mathbf{R}_{ij})]\,
 \mathbf{p}_{ij}\times\delta(\mathbf{r}_{ij})\,\mathbf{p}_{ij}\cdot
 (\mathbf{s}_i+\mathbf{s}_j)\,, \label{eq:DDLS}
\end{equation}
where $\mathbf{R}_{ij}=(\mathbf{r}_i+\mathbf{r}_j)/2$.
Because the $3N$ LS interaction is short range,
effects of the third nucleon are well approximated
by a linear function of $\rho$ for the coefficient $D[\rho]$.
The functional form of $D[\rho]$ is therefore taken to be
\begin{equation}
D[\rho(\mathbf{r})] = -w_1\,\frac{\rho(\mathbf{r})}
 {1+d_1\rho(\mathbf{r})}\,. \label{eq:DinLS}
\end{equation}
The $d_1$ term in the denominator suppresses instability in the MF
towards extremely high densities.
We adopt $d_1=1.0\,\mbox{fm}^3$,
after confirming that the results of the MF calculations
are insensitive to $d_1$.
The parameter $w_1$ is fixed so as to reproduce
the $n0i_{13/2}$-$n0i_{11/2}$ splitting with M3Y-P6 at $^{208}$Pb.
This new interaction is called M3Y-P6a.
Except for the LS interaction,
the other terms of M3Y-P6a are identical to the original M3Y-P6 interaction.

The coefficient $D[\rho]$ in $v^{(\mathrm{LS}\rho)}$ increases as the density grows,
and makes the $\ell s$ potential stronger (weaker)
in the nuclear interior (exterior).
Therefore the s.p. function of the $j=\ell+1/2$ ($j=\ell-1/2$) orbit
tends to shift inward (outward),
as has been confirmed in Fig.~1 of Ref.~\cite{ref:NI15}.
This mechanism gives a certain improvement
in the isotope shifts of the Pb nuclei.
It also plays a crucial role in reproducing the close charge radii
between $^{40}$Ca and $^{48}$Ca.
We shall illustrate these by the spherical HFB calculations with M3Y-P6a,
in comparison with those with M3Y-P6.
The spherical HFB results with D1S are also presented.
Together with M3Y-P6,
the D1S interaction will show the general trend that is given
by conventional interactions without the $3N$ LS term.

\subsection{Pb isotopes}
\label{subsec:Pb}

In the Pb nuclei,
it is customary to define the isotope shifts
by taking $^{208}$Pb as a reference,
$\mathit{\Delta}\langle r^2\rangle_p(\mbox{$^A$Pb})
=\langle r^2\rangle_p(\mbox{$^A$Pb})-\langle r^2\rangle_p(\mbox{$^{208}$Pb})$.
We depict $\mathit{\Delta}\langle r^2\rangle_p(\mbox{$^A$Pb})$
in Fig.~\ref{fig:Pb_drp},
comparing the HFB results with M3Y-P6a (red solid line)
to those with M3Y-P6 (green dashed line).

The lowest neutron s.p. level beyond $N=126$ is $n1g_{9/2}$~\cite{ref:TI}.
It is known that the partial $n0i_{11/2}$ occupation due to the pair correlation
increases the slope of the isotope shift
$\mathit{\Delta}\langle r^2\rangle_p(\mbox{$^A$Pb})$ at $N>126$.
Therefore the s.p. spacing between $n1g_{9/2}$ and $n0i_{11/2}$
is a quantity significant to the kink.
This s.p. spacing is largely influenced by the isospin content
of the LS interaction~\cite{ref:SLR94,ref:RF95}.
However, the reproduction of the kink only from the $2N$ LS interaction
requires that the $n1g_{9/2}$ and $n0i_{11/2}$ levels are almost degenerate,
or even inverted~\cite{ref:GSR13}.
This is in contradiction with the experimental data of the s.p. levels
extracted from the low-lying states of $^{209}$Pb~\cite{ref:TI}.
When the $3N$ LS interaction (or $\rho$-dependent LS interaction)
is considered,
the mean radius of $n0i_{11/2}$ increases and produces a stronger kink at $N=126$
as long as $n0i_{11/2}$ is populated to certain degree.
This effect of the $3N$ LS interaction on the isotope shifts
is confirmed by comparing the M3Y-P6a results to those of M3Y-P6.
With M3Y-P6a,
we have $\varepsilon(n0i_{11/2})-\varepsilon(n1g_{9/2})=0.72\,\mbox{MeV}$
at $^{208}$Pb,
close to the observed energy difference between $9/2^+$ and $11/2^+$
at $^{209}$Pb ($0.78\,\mbox{MeV}$).
We shall mention that the kink at $N=126$ is nevertheless reproduced,
comparably well to the models presented in Refs.~\cite{ref:SLR94,ref:RF95}.

It is remarked that both $v^{(\mathrm{LS})}$ and $v^{(\mathrm{LS}\rho)}$
are equally important for the description of the kink.
$v^{(\mathrm{LS})}$ is realistic and has reasonable isospin content,
giving appropriate s.p. spacing between $n1g_{9/2}$ and $n0i_{11/2}$.
If the s.p. spacing is too large
as it is the case with most Skyrme and Gogny interactions
that have a contact LS force,
pair excitations to $n0i_{11/2}$ are negligible
and the slope beyond $N>126$ is not described well,
even if $v^{(\mathrm{LS}\rho)}$ is introduced
and the s.p. function of $n0i_{11/2}$ distribute broadly.

In Ref.~\cite{ref:BBH06},
the generator-coordinate method was applied to the isotope shifts
of the Pb nuclei,
by using the SLy4 parameter-set of the Skyrme interaction.
While influence of correlations beyond the MF approximations
was observed in light Pb nuclei with $100\lesssim N\lesssim 115$,
related to the shape coexistence in this region,
it was found that the kink at $N=126$ could hardly be reproduced.

Let us stress that the LS part of M3Y-P6a ($v^{(\mathrm{LS})}+v^{(\mathrm{LS}\rho)}$)
is almost realistic,
since it is comprised of the $2N$ LS part of the M3Y-Paris interaction
and the $3N$ interaction suggested by the microscopic $\chi$EFT.
The $w_1$ parameter is, however, still a free parameter,
not well determined from the $\chi$EFT.
This is mostly because the $\chi$EFT is not yet a convergent framework.
Still, it is emphasized that the $\chi$EFT provides an important extension
of the LS interaction,
which gives a qualitative improvement for the description of isotope shifts
in Pb.

\subsection{Ca isotopes}
\label{subsec:Ca}

\begin{figure}
\resizebox{0.5\textwidth}{!}{%
  \includegraphics{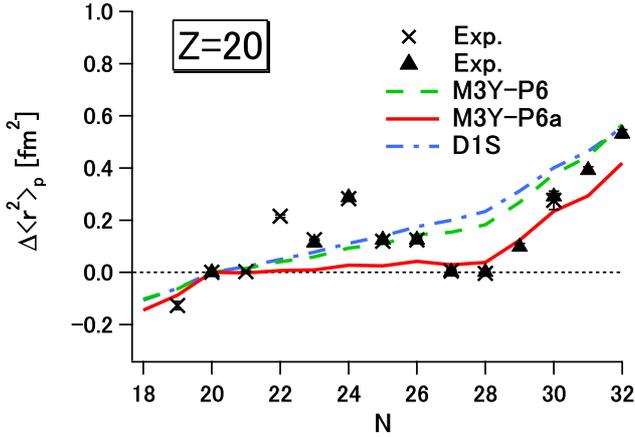}
}
\caption{Isotope shifts of the Ca nuclei
$\mathit{\Delta}\langle r^2\rangle_p(\mbox{$^A$Ca})$.
Experimental data are taken from Ref.~\protect\cite{ref:shift-Ca52} (triangles)
as well as from \protect\cite{ref:Ang13} (crosses).
See Fig.~\protect\ref{fig:Pb_drp} for other conventions.}
\label{fig:Ca_drp}
\end{figure}

By taking $^{40}$Ca as a reference,
the isotope shifts of the Ca nuclei are defined by
$\mathit{\Delta}\langle r^2\rangle_p(\mbox{$^A$Ca})
=\langle r^2\rangle_p(\mbox{$^A$Ca})-\langle r^2\rangle_p(\mbox{$^{40}$Ca})$.
The HFB results are depicted in Fig.~\ref{fig:Ca_drp},
in comparison with the experimental data.

If the $3N$ LS interaction is taken into account,
the $n0f_{7/2}$ function shifts inward,
which reduces the radius of $^{48}$Ca.
Because of this effect,
we have $\mathit{\Delta}\langle r^2\rangle_p(\mbox{$^{48}$Ca}) \approx 0$,
\textit{i.e.} $\langle r^2\rangle_p(\mbox{$^{40}$Ca}) \approx
\langle r^2\rangle_p(\mbox{$^{48}$Ca})$, in the M3Y-P6a result.
Thus the close radii between $^{40}$Ca and $^{48}$Ca
are well described in the self-consistent calculation for the first time.
More detailed analysis is given in Ref.~\cite{ref:Nak15}.

Very recently it is reported~\cite{ref:shift-Ca52} that
\textit{ab initio} calculations with the $\chi$EFT $2N+3N$ interaction
reproduce the close radii between $^{40}$Ca and $^{48}$Ca,
by using the coupled-cluster
and the similarity-renormalization-group methods.
Consistent with these \textit{ab initio} results,
the present MF results with the semi-realistic interaction
seem to clarify what is key to solve this long-standing problem
and that it is closely linked to the other problems;
the origin of the $\ell s$ splitting
and the kink in the isotope shifts of the Pb nuclei.

Having sizable deviation from those of $^{40,48}$Ca,
the experimental isotope shifts of $^{42-46}$Ca
are not described by the spherical HFB calculation.
This discrepancy will be ascribed to effects beyond MF,
which may include excitations out of the $^{40}$Ca core~\cite{ref:CLMN01}.


\subsection{Sn isotopes}
\label{subsec:Sn}

\begin{figure*}
\centerline{\resizebox{0.65\textwidth}{!}{%
  \includegraphics{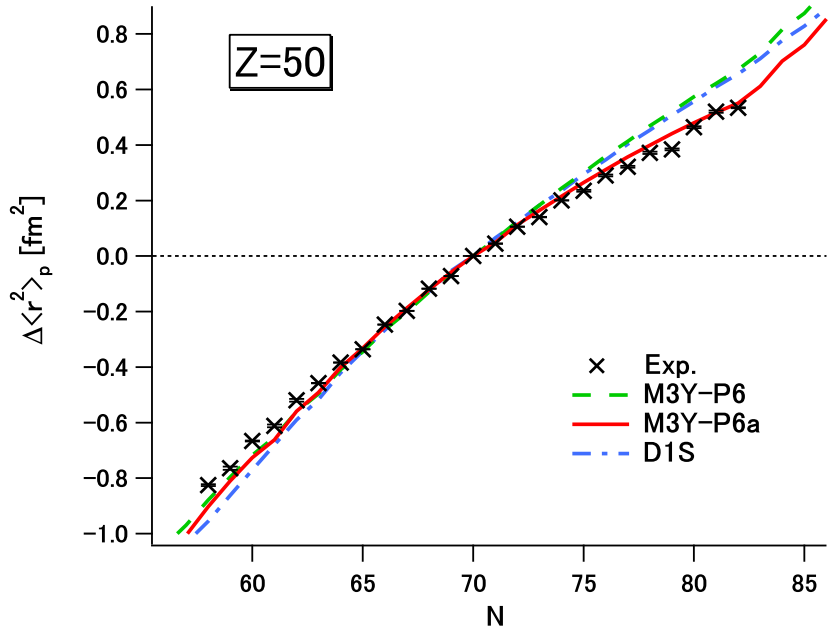}
}}
\caption{Isotope shifts of the Sn nuclei
$\mathit{\Delta}\langle r^2\rangle_p(\mbox{$^A$Sn})$.
See Fig.~\protect\ref{fig:Pb_drp} for conventions.}
\label{fig:Sn_drp}
\end{figure*}

We next present the isotope shifts of the Sn nuclei in Fig.~\ref{fig:Sn_drp},
$\mathit{\Delta}\langle r^2\rangle_p(\mbox{$^A$Sn})
=\langle r^2\rangle_p(\mbox{$^A$Sn})-\langle r^2\rangle_p(\mbox{$^{120}$Sn})$,
by adopting $^{120}$Sn as a reference.

The spherical HFB calculations with M3Y-P6a reproduce
$\mathit{\Delta}\langle r^2\rangle_p(\mbox{$^A$Sn})$
in a long chain of the Sn isotopes.
In particular, the curve in $70<N<82$ is in good agreement with the data,
in which the occupation of $n0h_{11/2}$ plays a significant role.
Since the $3N$ LS interaction shrinks the s.p. function of $n0h_{11/2}$,
the slope of $\mathit{\Delta}\langle r^2\rangle_p(\mbox{$^A$Sn})$
comes less steep in $70<N<82$.
Moreover, we predict a kink at $N=82$ in the M3Y-P6a result.
This takes place due to the pair excitation to $n0h_{9/2}$,
similarly to the kink at $N=126$ in the Pb nuclei.
Such a kink is not observed in the results without the $3N$ LS interaction,
even if correlations beyond MF are taken into account~\cite{ref:BBH06}.
Thus it is generally expected
that the $3N$ LS interaction tends to give a kink in the isotope shifts,
or to make a kink stronger, at the neutron $jj$-closed shell.
Measurements of $\mathit{\Delta}\langle r^2\rangle_p(\mbox{$^A$Sn})$
beyond $N=82$ are of great interest,
which may confirm the $3N$ LS effects further.

\section{Conclusion}
\label{sec:conclusion}

The M3Y-type semi-realistic interaction has been applied
to the self-consistent MF calculations under the spherical symmetry.
Particular focus is placed on roles of the non-central forces,
for which we refer to the realistic interaction.
The non-central forces on a realistic bases is significant
to describe nuclear shell structure reasonably well.
This seems to resonate Gogny's original idea
to determine effective central force.
Several examples have been presented:
(i) $p0d_{3/2}$-$p1s_{1/2}$ inversion in Ca isotopes,
to which the realistic tensor force gives correct $N$-dependence,
(ii) prediction of magic numbers in a wide range of the mass table,
to which the M3Y-P6 interaction gives results
compatible with almost all available data with a few exceptions,
(iii) isotope shifts of $Z=\mbox{magic}$ nuclei,
the kink at $N=126$ in Pb and close radii of $^{40,48}$Ca in particular,
which are greatly improved by the $3N$ LS interaction
indicated by the $\chi$EFT,
in addition to the realistic $2N$ LS interaction.
It is now clear that realistic interaction is definitely useful
for nuclear mean-field approaches,
and besides, semi-realistic interactions may bridge the gap
between \textit{ab initio} calculations and MF (or EDF) approaches.

\section*{Acknowledgments}
This work is financially supported in part
by JSPS KAKENHI Grant Number~24105008 and Grant Number~16K05342.

%
%

\end{document}